\documentclass[12pt]{article}
\usepackage{amssymb}
\usepackage{graphicx}
\usepackage{sint,cite}

\newcommand\be{\begin{equation}}
\newcommand\ee{\end{equation}}
\newcommand\bea{\begin{eqnarray}}
\newcommand\eea{\end{eqnarray}}
\newcommand\ba{\begin{array}}
\newcommand\ea{\end{array}}

\newcommand{\eq}[1]{Eq.~(\ref{#1})}
\newcommand{\fig}[1]{Fig.~\ref{#1}}
\newcommand{\tab}[1]{Table~\ref{#1}}
\newcommand{\sect}[1]{Section~\ref{#1}}
\newcommand{\ev}[1]{\left\langle #1 \right\rangle}
\def\MSb{\overline{{\rm MS}}}

\def\fm{\,{\rm fm}}
\def\Nf{N_{\rm f}}

\def\gqq{\bar{g}_\mathrm{qq}}
\def\gc{\bar{g}_\mathrm{c}}
\def\betac{\beta_\mathrm{c}}
\def\gms{\bar{g}_\mathrm{\overline{MS}}}
\def\bq{\bar{q}}
\def\rI{r_{\rm I}}
\def\rt{\tilde{r}}
\def\CF{C_{\rm F}}
\def\CA{C_{\rm A}}
\newcommand{\Estat}{E_{\rm stat}}
\newcommand{\Sshyp}{\mathcal{S}_{\rm sHYP}}

\def\taui{\tau_{\rm int}}
\def\Tr{{\rm Tr}\,}
\def\opsi{\overline{\psi}}
\def\bh{\bar{h}}
\def\oQ{\overline{Q}}
\def\rb{r_{\rm b}}

\def\as{\alpha_{\rm s}}
\def\ac{\alpha_{\rm c}}
\def\aqq{\alpha_{\rm qq}}
\def\bc{b^{\rm(c)}}
\def\bqq{b^{\rm(qq)}}
\def\gE{\gamma_{\rm E}}
\newcommand{\onecol}[2]{
        \begin{minipage}[t]{#1}{#2\vfill} \end{minipage}
        }

\def\ma[#1,#2,#3,#4]  {{\left( \matrix{ #1  & #2 \cr
                                        #3  & #4 \cr } \right)}}

\begin{document}

\thispagestyle{empty}
\title{{\normalsize
\mbox{} \hfill
\onecol{4.0cm}{\vspace{-1.9cm} DESY 10-172 \\
SFB/CPP-10-90 \\ WUB/10-24 \\ BUW-SC 2010/5}} \\
\vspace{1cm}
Determination of the Static Potential with Dynamical Fermions
}

\author{ 
\centerline{
           \includegraphics[width=2.5cm]{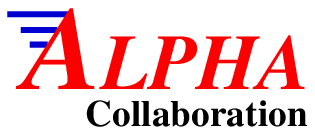}}\\\normalsize 
Michael Donnellan $^{a}$, Francesco Knechtli $^{b}$, Bj\"orn Leder $^{b,c}$, 
            Rainer Sommer $^{a}$ \\[0.5cm] \normalsize
$^{a}$ NIC, DESY, Platanenallee 6, D-15738 Zeuthen, Germany\\[0.25cm]\normalsize
$^{b}$ Department of Physics, Bergische Universit\"at Wuppertal\\\normalsize
Gaussstr. 20, D-42119 Wuppertal, Germany\\[0.25cm]\normalsize
$^{c}$ Department of Mathematics, Bergische Universit\"at Wuppertal\\\normalsize
Gaussstr. 20, D-42119 Wuppertal, Germany\\[0.5cm]
}
\date{}

\maketitle

\begin{abstract}
We present in detail a technique to extract the potential between a
static quark and anti-quark pair from Wilson loops measured on dynamical
configurations. This technique is based on HYP smearing and leads to an
exponential improvement of the noise-to-signal ratio of Wilson loops.
We explain why the correct continuum potential is obtained and show
numerical evidence that the cut-off effects are small. 
We present precise results for the non-perturbative potential. As
applications, we determine the scale $r_0/a$ and study the shape of the
static potential in the range of distances around $r_0$, where it can be
compared with phenomenological potential models.
\end{abstract}


\newpage

\section{Introduction}

The potential $V(r)$ between a static (infinitely massive) quark and
anti-quark pair separated by distance $r$ can be computed
from lattice quantum chromodynamics (QCD). It is extracted from the
expectation values of Wilson loops, which are traces of products of links
along rectangular paths extending in Euclidean time and 
one spatial direction.
In this article we consider only on-axis Wilson loops but off-axis
(non-planar) Wilson loops can also be used. 
Alternatively the static potential can
be extracted from the correlator of two Polyakov loops. Due to confinement,
the signal of Wilson loops falls exponentially with the area of the loop
(until string breaking sets in) but 
their variance is approximately constant.
In the statistical average of standard Monte Carlo lattice simulations,
the signal of Wilson loops is the result of strong
cancellations between positive and negative contributions.
This leads to an exponentially growing
noise-to-signal ratio which prevents the calculation of the potential at large
distances.

In pure gauge theory this problem has a cure. An exponential suppression of the
statistical noise of Wilson loops can be achieved by
the multi-hit (or one-link) method \cite{Parisi:1983hm} and much further by the 
multilevel algorithm \cite{Luscher:2001up}. These algorithms
are not applicable in presence of dynamical fermions due to
the non-locality of the effective gauge action when the logarithm of the
fermion determinant is included. In \cite{Hasenfratz:2001hp} a smearing
technique called hypercubic (HYP) smearing was introduced which can also be used
to measure Wilson loops in the presence of dynamical fermions
\cite{Hasenfratz:2001tw}. In pure gauge theory it was demonstrated 
in \cite{Alexandrou:2001ip} that the use of HYP smeared links
leads to a determination of the static potential comparable in precision to
the multi-hit method. In \cite{DellaMorte:2003mn} a new action for static
quarks was proposed, which uses HYP smeared links in the time covariant
derivative of the Eichten-Hill action. This leads to an exponential reduction
(compared with using the Eichten-Hill action) of the
noise-to-signal ratio for heavy-light correlation functions. This effect is
due to the fact that HYP smearing in the static action reduces the coefficient
of the divergent part of the self-energy of a static quark
\cite{DellaMorte:2005yc,Grimbach:2008uy}.

The interest in the determination of the static potential $V(r)$ through 
lattice simulations is twofold.
On the one side, there is the possibility to set the scale 
(i.e., determine the lattice spacing) through the scale $r_0$
introduced in \cite{Sommer:1993ce}.
The latter is defined from the static force $F(r)=V^\prime(r)$ as the solution
of
\bea
\left. r^2\,F(r)\right|_{r=r_0} & = & 1.65 \,. \label{r0}
\eea
A physical value for the scale $r_0\approx(0.45\ldots0.5)\fm$ can only be determined
through phenomenological potential models.
It is desirable for an absolute determination of the lattice spacing
to use a quantity which is directly accessible from experiment and replace
$r_0$ by a quantity like a hadron mass or decay constant.
But still, $r_0$ is very useful for a relative scale setting.

On the other side, the static potential is an interesting observable by itself
for phenomenology (see the conclusions) and to study the structure of gauge theories 
\cite{Maldacena:1998im,Forini:2010ek,Erickson:2000af,Pineda:2007kz}.
It exhibits clear effects of dynamical fermions, such as
string breaking \cite{Schilling:1999mv}, 
see the latest study in QCD \cite{Bali:2005fu} and high precision
studies with multilevel algorithms in other models \cite{Pepe:2010na}. 
In order to study the potential at the distances where the string breaks,
operators which dominantly create static-light meson pairs have to be included
in addition to the Wilson loops and we plan to do so in the future.
In this article we will concentrate on the determination of the static
potential at distances smaller than the string breaking distance
$\rb\approx3\,r_0$ \cite{Sommer:1994fr}.
We will study the quantity
\bea
c(r) & = & \frac{1}{2}\,r^3\,F^\prime(r) \,. \label{ccoeff}
\eea
It is a physical, renormalized quantity, which can be used to define
a running coupling. In \cite{Luscher:2002qv} $c(r)$ has been determined with
high precision in pure gauge theory using a multilevel technique. We will
compute it in this article for QCD with $\Nf=2$ flavors of quarks.

In section two we will describe our techniques to extract the static
potential from HYP smeared Wilson loops. We explain why this
procedure leads to a determination of the continuum static potential up to
$\mathcal{O}(a^2)$ lattice artifacts, which appear to be small. In section three we
present our results for the static potential, the scale $r_0/a$ and the
quantity $c(r)$ determined on a configuration ensemble generated with Wilson
gauge action and $\Nf=2$ flavors of $\mathcal{O}(a)$ improved Wilson quarks at
$\beta=5.3$. The quark mass corresponds to a pseudoscalar 
mass value close to $r_0\,m_{\rm PS}=1$ and we get a value $r_0/a=6.75(6)$.

\section{Techniques}

\subsection{Static potential with HYP smearing}

We measure $r/a \times T/a$ on-axis Wilson loops $W(r,T)$ 
on gauge configurations generated with $\Nf=2$ dynamical fermions. 
The technique is based on HYP smearing and was
introduced in \cite{Hasenfratz:2001hp}.
Before measuring the Wilson loops, we replace
{\em all} the gauge links by HYP-smeared ones.
We consider two choices of the HYP-smearing parameters:
one is
\be
        \alpha_1=0.75\,,\quad \alpha_2=0.6\,,\quad \alpha_3=0.3\,,
\label{hyp_par}
\ee
which we refer to as HYP, and the other is
\be
        \alpha_1=1.0\,,\quad \alpha_2=1.0\,,\quad \alpha_3=0.5\,,
\label{hyp2_par}
\ee
which we refer to as HYP2. 
We adopt the approximate projection onto $SU(3)$ as described 
in \cite{DellaMorte:2005yc}
and always use Eq. (2.24) and four iterations of Eq. (2.25) 
in \cite{DellaMorte:2005yc}.

In the following we show that this procedure leads
to a determination of the potential between quark and anti-quark sources
that agrees with the continuum potential up to $\mathcal{O}(a^2)$ effects
(after renormalization). The ingredients in this demonstration
are the selfadjoint positive transfer matrix of the lattice gauge theory
with Wilson fermions and Wilson plaquette action (rigorously proven
\cite{Luscher:1976ms}) as well as the existence and universality of
the continuum limit of the lattice theory with a static quark (lowest order
of heavy quark effective theory~\cite{Eichten:1989zv}).
The latter property is generally assumed and has been tested frequently (see 
\cite{Sommer:2010ic} for a longer discussion).
%
\begin{figure}[t]
 \begin{center}
     \includegraphics[width=10cm]{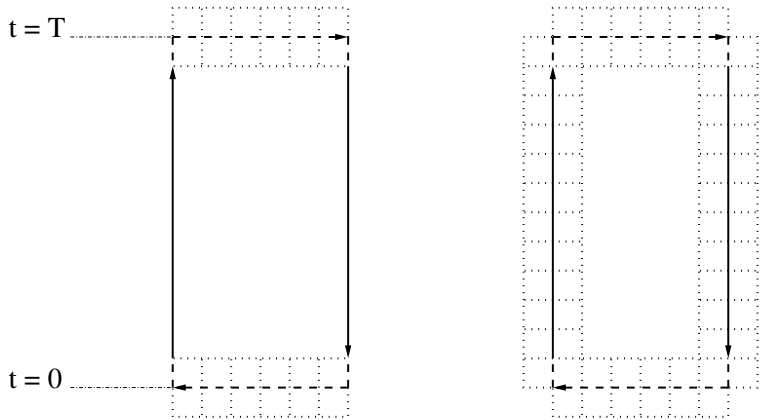}
 \end{center}
 \vspace{-0.5cm}
 \caption{\small Schematic representation of the measurement of Wilson loops.
In the first step (left figure) only the spatial Wilson lines are HYP-smeared: 
this corresponds to the definition of an operator $\hat{O}^\dagger$ 
that creates a $|Q\oQ(r)\rangle$ state.
In the second step (right figure) the temporal Wilson
lines are HYP-smeared: this corresponds to the choice of the static quark
action (and a modification of the operator $\hat{O}$).
}
 \label{f_hyppot}
\end{figure}
%

For the purpose of showing Eq.(\ref{Vartefacts}),
we split the HYP-smearing of the links used in building a Wilson loop into two steps,
which are schematically represented in \fig{f_hyppot}.
In the first step we consider Wilson loops where {\em only}
the space-like links are
HYP smeared. The smearing involves links at time-slices\footnote{ 
In total there are $N_t$ time-slices and periodic boundary
conditions are imposed in all directions.}
$t=-a,0,a$ and $t=T-a,T,T+a$ and corresponds in the Hamiltonian formalism to
an operator $\hat{O}^\dagger$ and $\hat{O}$ that creates or annihilates
a state $|\psi^{Q\oQ}(r)\rangle$ consisting of a static quark
and anti-quark pair at time-slices $t=a$ and $t=T-a$ respectively. The static
sources are separated
by a distance $r$ along one of the spatial directions.
The path integral average of this Wilson
loop can be written as a quantum mechanical expectation value
\be
\ev{W(r,T)} =
\frac{\Tr\left\{\mathbb{T}_0^{N_t-T/a-2}\hat{O}\mathbb{T}_{q\bq}(r)^{T/a-2}
\hat{O}^\dagger\right\}}
{\Tr\left\{\mathbb{T}_0^{N_t}\right\}} \,, \label{Wloop_RHYP}
\ee
where $\mathbb{T}_0$ is the transfer matrix in the gauge-invariant (or zero
charge) sector of the
Hilbert space, $\mathbb{T}_{q\bq}(r)$ the transfer matrix in the sector
with a quark and an anti-quark source separated by $r$ and $\Tr$ is the
operator trace in the Hilbert space. We denote the transfer matrix in the
temporal gauge (where the time-like links are set to identity)
by $\mathbb{T}_{\rm temp}$. The Hamiltonian $\mathbb{H}$ is defined through
$a\mathbb{H}=-\ln\{\mathbb{T}_{\rm temp}\}$.
For the theory with Wilson quarks without a clover term $\mathbb{T}_{\rm temp}$
has been constructed in \cite{Luscher:1976ms}. 
The transfer matrix in a specific charge sector is
obtained by multiplying $\mathbb{T}_{\rm temp}$ with the projectors onto that
charge sector. Note that the representation \eq{Wloop_RHYP} differs from the usual
one only in that the operators $\hat{O}$ represent fields in the path integral
on three timeslices, not one. If written down explicitly in the form 
of \cite{Luscher:1976ms} they involve integration kernels. But their 
explicit form is not relevant here. 
Using the spectral decomposition of the transfer matrices
(see for example \cite{Knechtli:1999tw}) and
taking the limit $N_t\to\infty$, \eq{Wloop_RHYP} becomes
\bea
\ev{W(r,T)} & \stackrel{N_t\to\infty}{\sim} &
\sum_n c_nc_n^* {\rm e}^{-V_n(r)(T-2a)}\,, \label{Wloop_RHYP_spectral}
\eea
where $c_n=\langle n\,;q\bq|\hat{O}^\dagger|0 \rangle$ are the overlaps of states
obtained by applying the operator $\hat{O}^\dagger$ to the vacuum $|0 \rangle$
with the eigenstates $|n\,;q\bq\rangle$ of the Hamiltonian 
(with eigenvalues $E^{(q\bq)}_n(r)$)
in the sector with a quark and an anti-quark source. 
In \eq{Wloop_RHYP_spectral} we use $V_n(r)=E^{(q\bq)}_n(r)-E_0^{(0)}$,
where $E_0^{(0)}$ is the vacuum energy. 
For example $V_0(r)$ is the static potential and 
$V_1(r)$ its first excitation.

In the second step we rewrite the Wilson loop as a path integral expectation
value
\bea
\ev{W(r,T)} & = & -{1\over2}
\left\langle \opsi_h(a,\vec{0})P_-(a,\vec{0};a,r\hat{k})\gamma_5\psi_{\bh}(a,r\hat{k})
\right.\nonumber \\
&& \left.
\opsi_{\bh}(T-a,r\hat{k})P_+^\dagger(T-a,\vec{0};T-a,r\hat{k})\gamma_5\psi_h(T-a,\vec{0})
\right\rangle \,, \label{Wloop_pathi}
\eea
where $\psi_h,\,\opsi_h$ and $\psi_{\bh},\,\opsi_{\bh}$ are the
static quark and anti-quark fermion fields respectively\footnote{
The prefactor $-\frac12$ and the gamma-matrices are due to our choice
of treating the static quark fields as 2-component static fermion fields,
see for example \cite{Sommer:2010ic}.} 
and $P_\pm(t,\vec{0};t,r\hat{k})$ 
represents the gauge parallel transporter made from a product of space-like HYP-links
at time $t\pm a$ and temporal links at time $t\pm a$ in $\mp$-direction 
(dashed lines in \fig{f_hyppot}).

After integration over the static fields,
the static quark propagator generates the time-like links in the observable, 
cf. Eq. (2.4) in \cite{DellaMorte:2005yc}, and
one recovers the Wilson loops. Different choices for the static quark action can be made,
in particular we consider here the one where the covariant derivative in time
involves HYP or HYP2 links\footnote{
We smear also the temporal links contained in the definition of the parallel
transporters $P_\pm$ in \eq{Wloop_pathi}. This corresponds to a change in the definition
of the operator $\hat{O}$ in \eq{Wloop_RHYP} and has no consequences for the argument
presented here.} \cite{DellaMorte:2003mn}. It was shown in
\cite{Necco:2001xg} that static potential differences (where the self energy
is canceled) have $\mathcal{O}(a^2)$ leading lattice artifacts, essentially
due to the automatic $\mathcal{O}(a)$ improvement of heavy quark effective
theory \cite{Kurth:2000ki}. 
This is true in the theory with dynamical
fermions provided that they are $\mathcal{O}(a)$ improved.\footnote{For 
Wilson fermions
improvement is achieved by adding the clover
term~\cite{impr:SW,impr:pap1,impr:csw_nf2} or by using a twisted mass 
term~\cite{tmqcd:pap1}
``at maximal twist''\cite{tmqcd:FR1}.} 
We therefore conclude
\be
V_n^{\rm HYP/HYP2}(r) -2\Estat^{\rm HYP/HYP2}
 = V_n^{\rm continuum}(r) -2\Estat^{\rm continuum}+ \mathcal{O}(a^2)
\,, \label{Vartefacts}
\ee
%
\begin{figure}[t]
 \begin{center}
     \includegraphics[width=7cm]{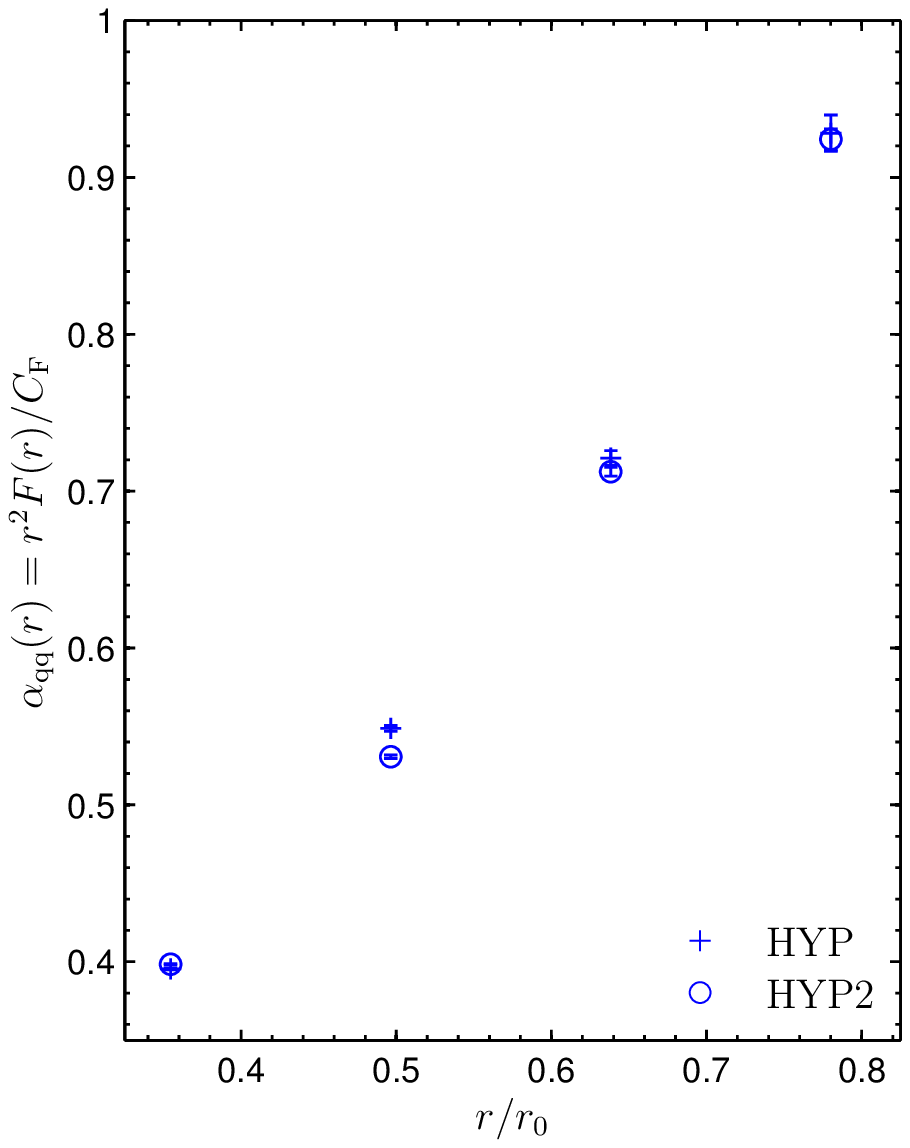}
     \hfill
     \includegraphics[width=7cm]{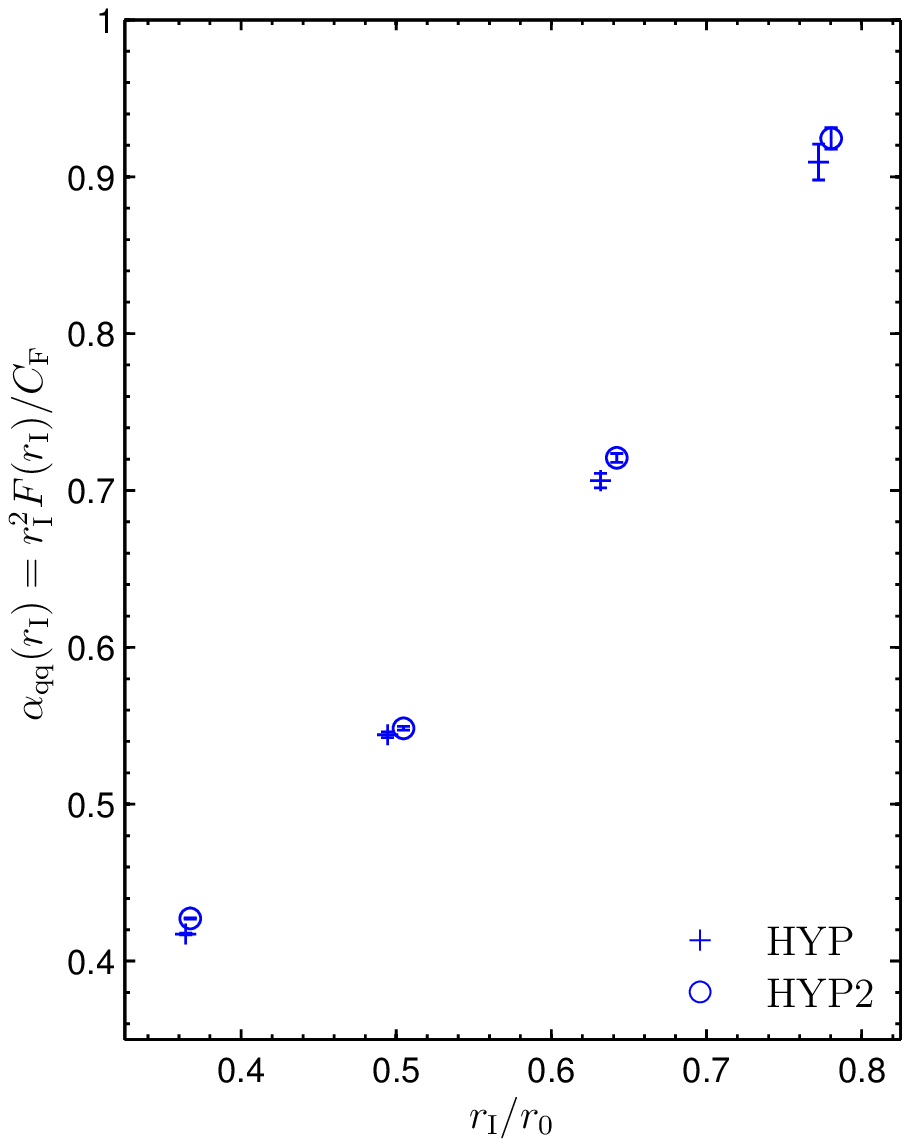}
 \end{center}
 \vspace{-0.5cm}
 \caption{\small The qq-coupling $\alpha_{\mathrm{qq}}(r=1/\mu)$ obtained from the static force
   $F(r-a/2)  =  [V(r)-V(r-a)]/a$ with two different choices of the static action (left panel). In
   \sect{s_force} we give an improved definition of the force which is free of cutoff effects at
   tree level of perturbation theory (right panel).}
 \label{f_aqq_HYP_vs_HYP2}
\end{figure}
%
where for convenience we have subtracted $V(\infty)=2\Estat$. Here
$\Estat$ is the  binding energy of a meson made
of a static and a light dynamical quark.

In order to investigate the magnitude of the lattice artifacts we compare
in the left panel of \fig{f_aqq_HYP_vs_HYP2} the qq-coupling
 $\alpha_{\mathrm{qq}}(r)=r^2 F(r) /C_\mathrm{F}$
for the HYP and HYP2 actions. The static force $F(r)$ is obtained from the static potential
$F(r-a/2) = [V(r)-V(r-a)]/a$. Details about the extraction
of the static potential from correlation functions of Wilson loops are presented
in the next section. We use the dynamical configurations described in \sect{s_res}.
The difference of the couplings is given, to leading order in the cut-off effects, by
$a^2/r^2\, G(\Lambda r,mr)$. The function $G$ describes the
$r$-dependence and quark-mass-dependence of the cut-off effects. 
The size of the cut-off effects is small but with our errors 
they are significantly different from zero for $r<r_0$.
They happen to be most significant at $r\approx r_0/2$.

In \sect{s_force} we describe how improved observables can be defined
such that these cut-off effects are eliminated at tree
level and are substantially reduced non-perturbatively \cite{Necco:2001xg}.
The right panel of \fig{f_aqq_HYP_vs_HYP2} is the same as the left but
using the improved definition of the force \eq{Flat}. 
The cut-off effects are visibly reduced. We emphasize
that the figure is not sufficient to exclude cutoff effects
which are independent of the choice of static action. Different
lattice spacings are needed to study those.

Our choice of the static quark action and the smearing of the spatial links
is with parameters HYP2 \eq{hyp2_par}.
It gives a static potential with a somewhat better statistical precision than
with parameters HYP. This can be understood in perturbation theory: 
the HYP2 parameters are such that they approximately 
minimize the one-loop coefficient of the $1/a$ self-energy contribution
of a static quark \cite{DellaMorte:2005yc,Grimbach:2008uy}.  
Our data show that this property remains true non-perturbatively: 
we find $V^{\rm HYP2}-V^{\rm HYP}\approx -0.07/a \approx 
2(\Estat^{\rm HYP2}-\Estat^{\rm HYP})$.
For the last statement we use the results for
$\Estat$ of reference \cite{lat10:hqet}.

\subsection{Variational basis}

On the HYP2-smeared gauge link configurations $\{U(x,\mu)\}$, we measure
a correlation matrix of on-axis Wilson loops at fixed spatial extension 
$r/a$ and temporal extension $T/a$:
\bea
 C_{lm}(T) & = & \left<\mathrm{tr}\left\{P^{(l)}(0;r\hat{k})\, 
 P(r\hat{k};r\hat{k}+T\hat{0})\,
 P^{(m)\dagger}(T\hat{0};r\hat{k}+T\hat{0})
 P^\dagger(0,T\hat{0})\right\}\right> \,, \nonumber \\[-1ex]
 \label{Wloop_matrix}
\eea
where $P(x,y)$ represents the product of links connecting $y$ to $x$. Neglecting
the superscripts on the spatial $P$'s, Eq. (\ref{Wloop_matrix}) is equivalent to
Eq. (\ref{Wloop_pathi}) after integrating out the static fields. In the
product of spatial links, the superscript $P^{(l)}$ means that the links
$U_l(x,k)$ used in the product are obtained by applying the spatial smearing
$\Sshyp$ operator $n_l$ times
\bea
 U_l(x,k) & = & (\Sshyp)^{n_l}\,U(x,k) \,.
\eea
$\Sshyp$ means smearing with only two levels of HYP blocking with staples 
restricted to spatial directions 
and therefore it needs two parameters, which we set to $\alpha_2=0.6$
and $\alpha_3=0.3$. In the argument of the previous section the fat parallel
transporters $P^{(l)}$ correspond to operators $\hat{O}_l$ implementing trial states
$|\psi_l^{Q\oQ}(r)\rangle = \hat{O}_l^\dagger |0\rangle$.
In \cite{Guagnelli:1998ud} a formula for suitable smearing
parameters $n_l$ is given in the case of APE smearing. In order to choose $n_l$ 
for spatial HYP smearing, we use the result of
\cite{Bernard:1999kc}, that the mean squared extension of APE smearing is
approximately $\alpha_{\rm APE}\, n_{l,{\rm APE}}\, a^2/3$, and require that this
is equal to $n_l a^2$ for HYP smearing.
We get an approximate formula for a good range of HYP smearing levels
\bea
n_l & \approx & \frac{l}{12} \left(\frac{r_0}{a}\right)^2 \,.
\eea
For our data on the configuration ensemble E5g (see \sect{s_res})
we have computed a large correlation matrix using smearing levels 
$n_{0,1,2,3,5}$.
We find that this basis can be reduced to an optimal subset of $M=3$ levels
\bea
n_2=8\,,\quad n_3=12 \,,\quad n_5=20\,,\quad \mbox{at $\beta=5.3$} \,.
\eea
The higher smearing levels improve the determination of the energy levels.
%
\begin{figure}[!t]
 \begin{center}
     \includegraphics[width=11cm]{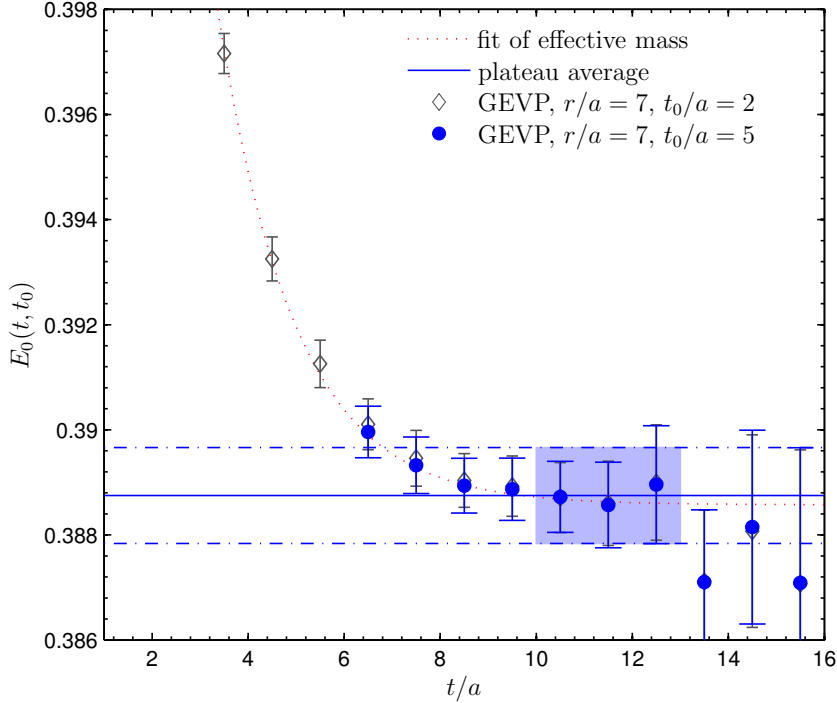}
 \end{center}
\vspace{-0.5cm}
\caption{\small Effective masses $E_0(t,t_0)$ (filled blue circles for
  $t_0/a=5$, empty black diamonds for $t_0/a=2$)
  for the ground state potential at $r=7a$. 
  The red dotted line is the fit \eq{fitgevp}. The blue
  line is the plateau average from the points in the blue
  shaded area (the blue dashed-dotted lines are the plateau errors).}
\label{f_E0}
\end{figure}
%

We use the generalized eigenvalue method
\cite{Campbell:1987nv,Luscher:1990ck,Guagnelli:1998ud,Blossier:2009kd} to
extract the ground state potential as follows. We first solve
the generalized eigenvalue problem
\bea
C(t)\,\psi_\alpha & = & \lambda_\alpha(t,t_0)\,C(t_0)\,\psi_\alpha \,.
\eea
Then we perform a fit to
\bea
E_\alpha(t+{a\over2},t_0) \equiv 
\ln\left(\lambda_\alpha(t,t_0)/\lambda_\alpha(t+a,t_0)\right) =
 E_\alpha + \beta_\alpha
{\rm e}^{-(E_M-E_\alpha)(t+{a\over2})} \,, \label{fitgevp}
\eea
with fit parameters $E_\alpha$, $\beta_\alpha$ and $E_M$,
simultaneously for $\alpha=0,1$ ($\alpha=0$ corresponds to the ground state,
$\alpha=1$ to the first excited state), $t_0/a=2,3,4$ and 
$t_0+a\le t\le 2t_0$ (the latter constraint is necessary for 
eq.(\ref{fitgevp}) to hold~\cite{Blossier:2009kd}
), 
i.e., we have 18 data points $E_\alpha(t+{a\over2},t_0)$ for 5 fit parameters. 
%
\begin{figure}[!t]
 \begin{center}
     \includegraphics[width=11cm]{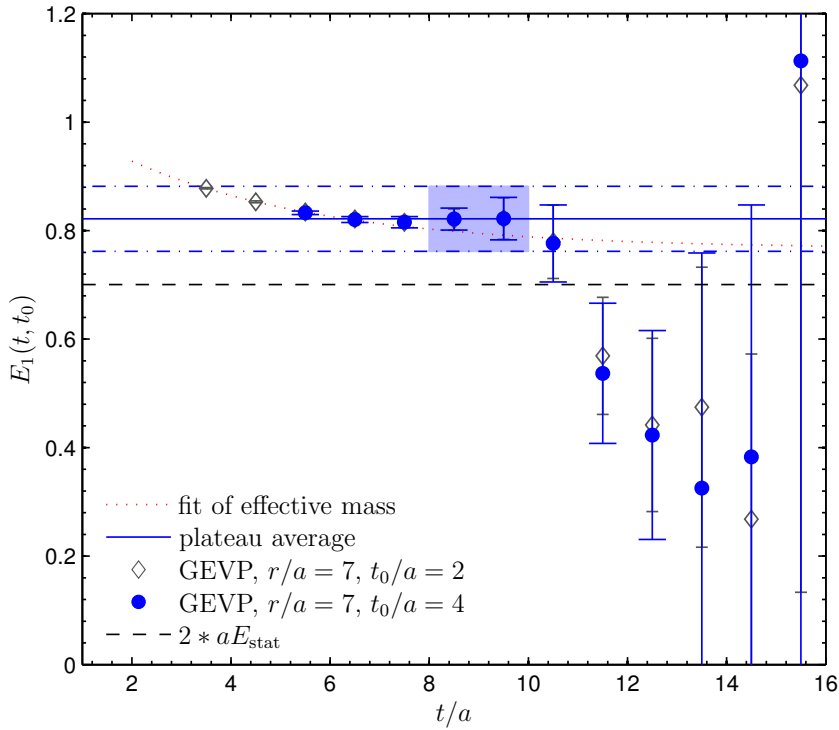}
 \end{center}
\vspace{-0.5cm}
\caption{\small Effective masses $E_1(t,t_0)$ (filled blue circles for
  $t_0/a=4$, empty black diamonds for $t_0/a=2$) for the first excited state
  potential at $r=7a$. The dashed black line represents the value 
  $2aE_{\rm stat}$, the meaning of the other curves is as explained in 
  \fig{f_E0}.}
\label{f_E1}
\end{figure}
%

The values of the ground state potential $V(r)$ as a function of $r$
are determined from a plateau average of the corresponding effective masses 
$E_0(t,t_0)$ starting at the value $t=2t_0+{a\over2}$, 
where the fixed value $t_0$ is determined by the requirement that
\bea
\sigma_{\rm sys}(E_0(2t_0+{a\over2},t_0)) \equiv 
\beta_0{\rm e}^{-(E_M-E_0)(2t_0+{a\over2})} \lesssim 
\frac{1}{4}\sigma_{\rm stat}(E_0(2t_0+{a\over2},t_0)) \,, \label{E1cond}
\eea
where $\sigma_{\rm sys}(\cdot)$ and $\sigma_{\rm stat}(\cdot)$ denote the 
systematic and statistical error respectively. For our data,
\eq{E1cond} is satisfied for $t_0/a=5$ for all values of $r$.
The plateau average is stopped before the time,
when either the difference of the effective mass with the one at
$t=2t_0+{a\over2}$ is larger than the statistical error of the latter or
the statistical error of the effective mass is larger than twice the one
of the effective mass at $t=2t_0+{a\over2}$.
The effective masses $E_0(t,t_0)$ (filled blue circles) together with
the fit \eq{fitgevp} (red dotted line) and the plateau average (blue
line with error band marked by blue dashed-dotted lines) for
$r=7a \approx r_0$ are shown in \fig{f_E0}. The plateau average comprises 
three points at $t/a=10.5,11.5,12.5$.
The error of the plateau average is the sum
of the statistical and the systematic errors, with  
the latter being given by the left-hand side
of \eq{E1cond}.
For comparison, we also plot in \fig{f_E0} the effective masses obtained using
$t_0/a=2$ (empty black diamonds). They are part of the data set fitted using
\eq{fitgevp}. At the times when they are both defined, the effective masses
for $t_0/a=5$ and $t_0/a=2$ agree with each other, which is somewhat
surprising since non-leading corrections certainly have a dependence 
on $t_0$. 

In principle the excited potentials can be determined in the same way. 
However, the analysis is complicated by
the dynamics of string breaking. From model studies
\cite{Philipsen:1998de,Knechtli:1998gf,Knechtli:2000df}
as well as from \cite{Bali:2005fu}, we know that an extraction of 
the potentials requires the inclusion of operators which dominantly create 
static-light meson pairs in addition to the string-like operators we use
here. Only then does the ground state at large distances $r>r_{\rm b}$ 
contribute significantly to the spectral decomposition of the correlation
function matrices at the accessible time separations
(cf. \cite{Gliozzi:2004cs}). 
While we are not concerned here with this string breaking region,
it is known \cite{Knechtli:2000df,Bali:2005fu} 
that for $r<r_{\rm b}$ the first excited state is
an (approximate) meson-anti-meson state at $V_1 \approx 2 E_{\rm stat}$.
This state is not well seen in our computation which does not 
include the meson pair operators. In \fig{f_E1} we show the effective masses
$E_1$ for $r=7a \approx r_0$. The dashed black line represents
$2aE_{\rm stat}=0.7007(14)$, the meaning of the other curves is as 
explained for \fig{f_E0}. Although the effective masses seem to form a 
plateau at times $t=8.5a$ and $t=9.5a$, they drop at values $t \geq 10.5a$, 
but the statistical errors are too large in this region to determine an 
energy level. For $r<r_0$ we see plateaus for $E_1$ which are compatible with 
$2aE_{\rm stat}$.

This deficiency of our variational basis also affects the estimate of the 
ground state potential, but here the only concern is the description of the 
corrections to the asymptotic plateau of the form $\beta_0{\rm e}^{-(E_M-E_0)t}$.
These enter the final numbers and errors only  
to estimate where we start the plateau average such that 
excited state contaminations are small compared to our statistical error.
A very precise determination of $E_M$ or in general of the excited states
is not necessary for this purpose. Furthermore, the effective mass 
figures show that our plateau selection is 
rather conservative; the extracted ground state potential
is reliable within the cited errors.

\subsection{Tree level improved force \label{s_force}}

In order to determine the scale $r_0$ \cite{Sommer:1993ce} from \eq{r0}
we will need the static force $F(r)$. An improved definition of the force
on the lattice is \cite{Sommer:1993ce,Guagnelli:1998ud,Necco:2001xg}
\bea
F(\rI) & = & [V(r)-V(r-a)]/a \,, \label{Flat}
\eea
where $\rI=r-a/2+\mathcal{O}(a^2)$ is chosen such that at tree level in
perturbation theory \cite{Weisz:1982zw} one has
\bea
F_{\rm tree}(\rI) = \CF\,\frac{g_0^2}{4\pi\rI^2} \,, \label{rIdef}
\eea
where $\CF=4/3$ for gauge group $SU(3)$.
The formula for $\rI$ depends on the static quark action and it is given in
Appendix \ref{s_appa} for HYP actions.
   
An improved lattice definition of the quantity $c(r)$ in \eq{ccoeff}
is given by \cite{Luscher:2002qv}
\bea
c(\rt) & = & \frac{1}{2}\rt^3[V(r+a)+V(r-a)-2V(r)]/a^2 \,, \label{clat}
\eea
where $\rt=r+\mathcal{O}(a^2)$ is chosen such that
\bea
c_{\rm tree}(\rt) & = & -\CF\,\frac{g_0^2}{4\pi}\,. \label{rtdef}
\eea
The formula for $\rt$ depends on the static quark action and it is given in
Appendix \ref{s_appa} for the HYP actions.
In Appendix \ref{s_appb} we give the 4-loop beta function for
the coupling $\ac=-c(r)/\CF$ which we will use to generate perturbative curves
for $c(r)$ to be compared with the lattice data.

\section{Results \label{s_res}}

We compute the static potential on the lattice ensemble E5g generated by the CLS
(``Coordinated Lattice Simulations'') project\footnote{
{\tt https://twiki.cern.ch/twiki/bin/view/CLS/WebHome}} at $\beta=5.3$,
$\kappa=0.13625$ with geometry $64\times32^3$ and periodic boundary conditions
for all fields apart from anti-periodic boundary conditions for the fermions in
time. 
The value of the pseudo-scalar mass is $am_{\rm PS}=0.15$.
The algorithm used in CLS is the
deflation accelerated DD-HMC algorithm \cite{Luscher:2005rx,Luscher:2007es}.
The trajectory length is $\tau=4$ and we separate the measurements of Wilson
loops by 4 trajectories. Given the block size $8^4$, the active links
represent 37\% of all links. Hence the separation in molecular dynamics time
between measurements is approximately 6 units (when all links are changed). 
We have a statistics of about 1000 measurements.
%
\begin{figure}[t]
 \begin{center}
     \includegraphics[width=11cm]{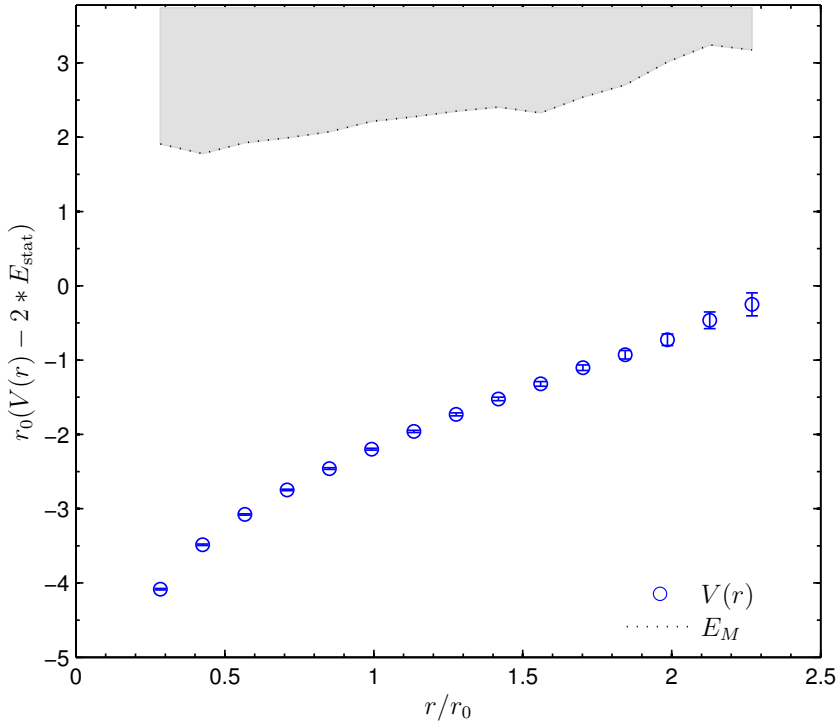}
 \end{center}
 \vspace{-0.5cm}
 \caption{\small Ground state potential $V(r)$ (circles). 
   The shaded area marks energy states larger than $E_M$
   (here $M=3$) determined from \eq{fitgevp}.
}
 \label{f_V}
\end{figure}
%

In \fig{f_V} we show the ground state potential $V$,  and for illustration
the rough estimate of the excitation $E_M$ (here $M=3$) in \eq{fitgevp}.
In order to get
renormalized quantities we subtract twice the binding energy $\Estat$
of a meson made of a static and a light dynamical quark.
Everything is made dimensionless by
appropriate powers of $r_0$ extrapolated to the chiral limit, see below. 
The first excited state potential is not shown due to the difficulties
described above. 

The range of string breaking is not yet reached.
We can estimate it from the condition $V(\rb)=2\,\Estat$ to be
\bea
2.4\;\le & \left.{\rb \over r_0}\right|_{r_0\,m_{\rm PS}=1.0} & \le\;2.6 \,.
\eea
For comparison, in \cite{Bali:2005fu} $\rb/r_0\approx2.5$ was found at a
larger quark mass corresponding to $r_0m_{\rm PS}=1.7$, albeit in the theory
without $\mathcal{O}(a)$ improvement.

%
\begin{table}[t]
 \centering
\begin{tabular}{clcl}
\hline\\
  $\rI/a$  & $a^2 F(\rI)$ & $\tilde{r}/a$ & $c(\tilde{r})$ \\[1.0ex] \hline\\
  3.55805 & 0.05776(12) &  4.046306 &-0.3596(41)\\
  4.52674 & 0.04690(19) &  5.026094 &-0.391(15) \\
  5.50073 & 0.04074(35) &  5.999703 &-0.405(41) \\
  6.48362 & 0.03699(48) &  6.977869 &-0.519(99) \\
  7.47397 & 0.03393(75) &  7.917429 &-0.35(22)  \\
  8.46922 & 0.0325(12)  &  &\\
  9.46734 & 0.0295(13)  &  &\\
 10.4670 & 0.0287(19)  &  &\\
 11.4676 & 0.0310(25)  &  &\\
 12.4685 & 0.0248(53)  &  &\\
 13.4697 & 0.0285(48)  &  &\\
 14.4709 & 0.0373(75)  &  &\\
 15.4721 & 0.030(11)   &  &\\[1.0ex]
\hline
\end{tabular}
 \caption{\small The values of the force $F(r)$ in lattice units and the
   physical quantity $c(r)$ at the accessible improved distances $\rI$ and
   $\tilde{r}$ respectively. We do not include values that require the
   potential $V(r=2a)$, since it may be affected by large cut-off effects. 
}
 \label{t_force_c}
\end{table}
%

The scale $r_0$ is defined from the condition \eq{r0}.
The static force is computed from \eq{Flat} using the improved
distance $\rI$ in \eq{rIeq}. In \mbox{Table \ref{t_force_c}} we list the values
of the force in lattice units. We do not include the force at
$\rI/a=2.58875$ because it requires the potential at distance $r=2a$, which
may be affected by relatively large cut-off effects.
We determine the solution of \eq{r0} by interpolation of the force $F$,
using a 2-point interpolation $F(r)=f_0+f_2/r^2$ and a 3-point
interpolation adding a $f_4/r^4$ term to control the systematic error (it is found 
to be negligible). We obtain
\bea
\left.\frac{r_0}{a}\right|_{am_{\rm PS}=0.15}& = & 6.747(59) 
\qquad (\beta=5.3) \,.
\eea
The error is determined by taking the upper bound $\taui=6$ (see below) and 
neglecting the systematic error (due to excited state contributions), 
which is much less than
the statistical one (due to condition Eq. (\ref{E1cond})).
In \cite{bjoernlat10} we presented a preliminary value $r_0/a=7.05(3)$
extrapolated to the chiral limit. Throughout this article we use this value
(without errors) for the purpose of plotting dimensionless quantities.
%
\begin{figure}[t]
 \begin{center}
     \includegraphics[width=11cm]{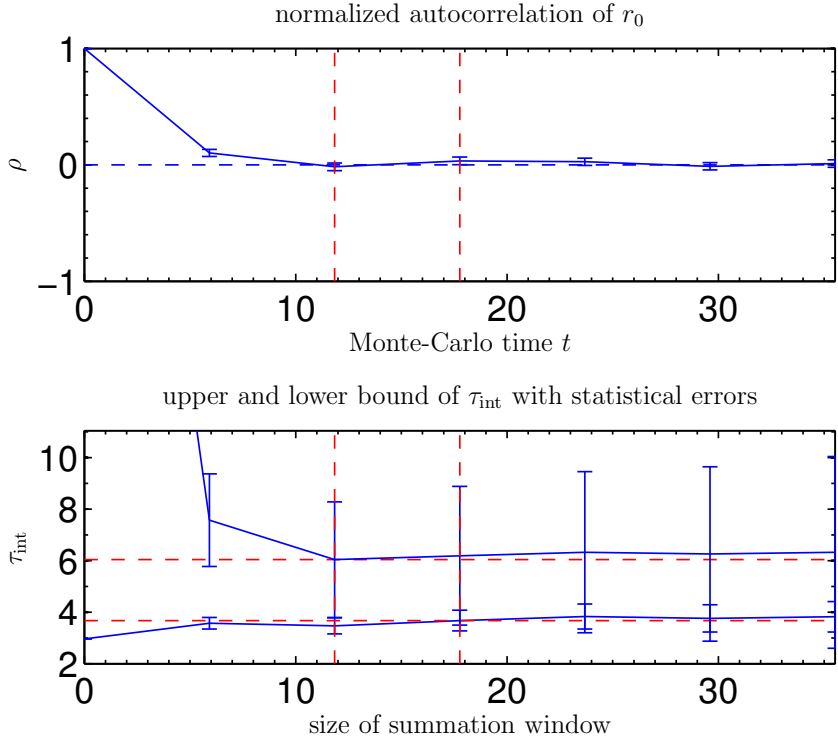}
 \end{center}
 \vspace{-0.5cm}
 \caption{\small Auto-correlation function $\rho(t)$ and integrated
   auto-correlation time $\taui$ of $r_0$. 
   The Monte-Carlo time is in units of molecular dynamics time.}
 \label{f_auto_r0}
\end{figure}
%

In \cite{Schaefer:2010hu} it was shown that the auto-correlation time of the
topological charge suffers from critical slowing down proportional to 
$a^{-5}$ in the present range of lattice spacings. However,
in the same reference it was shown that Wilson loops are affected by
a much milder critical slowing down, implying that their coupling to the slow
modes of the Monte Carlo simulation is small. A method for correcting the
error analysis, by adding a tail to the auto-correlation function that takes
into account the coupling to slow modes, was presented in
\cite{Schaefer:2010hu}. We use this method in our data analysis and we set
$\tau_{\rm exp}=39$ in molecular dynamics units\footnote{
In these units, the DD-HMC molecular dynamics time is multiplied by the ratio
of active links, which in our case is 37\%.}
from Table 4 of \cite{Schaefer:2010hu}. 
In \fig{f_auto_r0}
we show the auto-correlation function $\rho(t)$ and the
integrated autocorrelation time $\taui$ of $r_0$, 
determined with the program\footnote{{\tt http://www-zeuthen.desy.de/alpha/}} 
of \cite{Schaefer:2010hu}
implementing the method of \cite{Wolff:2003sm,Schaefer:2010hu}. 
The vertical dashed lines in the plots mark the applied
summation windows, the lower one
is used when we add the tail due to the slow modes, while 
the larger one comes from using the method of \cite{Wolff:2003sm}. 
Adding to the summed autocorrelation function the correction due 
to the slow modes leads to
the upper curve and upper bound on $\taui$, which we take for all
quantities in our analysis. The lower curve corresponds to $\taui$ determined
from \cite{Wolff:2003sm}.
For $r_0$ we get an upper bound $\taui=6$
which is a factor 6.5 smaller than $\tau_{\rm exp}$, but
a factor 1.5 larger than without accounting for effects of undetected
slow modes (lower bound).
%
\begin{figure}[t]
 \begin{center}
     \includegraphics[width=11cm]{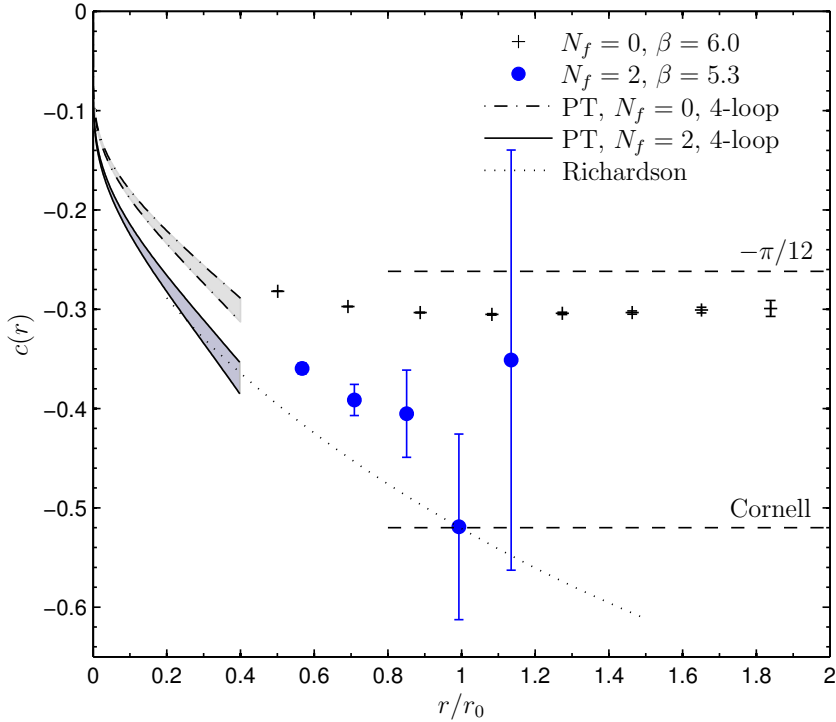}
 \end{center}
 \vspace{-0.5cm}
 \caption{\small The physical quantity $c(r)$ in \eq{ccoeff}. 
Comparison of $\Nf=2$ (circles)
   with $\Nf=0$ (pluses) Monte Carlo data (for $\Nf=0$ taken from
   \cite{Luscher:2002qv}) and perturbation theory. Also the value $c=-0.52$ in
the Cornell \cite{Eichten:1979ms} potential and the curve (dotted) derived
from the Richardson \cite{Richardson:1978bt} potential are plotted.
}
 \label{f_c}
\end{figure}
%

In \fig{f_c} we plot our result for the physical quantity  
$c(r)$ computed from \eq{clat} using the improved distance $\rt$ in \eq{rteq}.
The numbers are given in Table \ref{t_force_c}.
In order to compare our $\Nf=2$ results (circles), we plot them
together with the $\Nf=0$ data (pluses) of \cite{Luscher:2002qv} and with the
perturbative curves obtained using the 4-loop beta function (continued line for
$\Nf=2$, dashed-dotted line for $\Nf=0$). The perturbative formula for $c(r)$
is presented in Appendix \ref{s_appb} and we used the preliminary updated
value of the $\Lambda$ parameter presented in \cite{bjoernlat10}.
The spread of the perturbative curve reflects the uncertainty of the $\Lambda$
parameter. For a comparison with our Monte Carlo data, it is legitimate to
plot the perturbative curve of $c(r)$ in massless perturbation theory,
since quark mass corrections are of order $\alpha^2\times(m_qr)^2$ and 
are expected to be negligible at our small quark mass. 
The distances in \fig{f_c} are
normalized by $r_0$ extrapolated to the chiral limit. As the perturbative
curves already indicate, the value of $c$ for $\Nf=2$ is found to be
lower than for $\Nf=0$. 
In pure gauge theory,  $c(r)$ starts approaching the asymptotic
value $c(\infty)=-\pi/12$ with corrections of order $1/r$ as 
predicted from the effective bosonic string
theory \cite{Luscher:1980fr,Luscher:1980ac}. Our data for $\Nf=2$
have quite large errors when $r/r_0\ge1$. We compare them with the value
$c=-0.52$ that it takes in the phenomenological Cornell potential
\cite{Eichten:1979ms} and with the curve obtained from the Richardson
potential \cite{Richardson:1978bt}. Our data seem to follow the Richardson
curve for $r\lesssim r_0$ quite closely. 
It is not yet possible to tell whether there is a
plateau region around or above $r_0$ before string breaking sets in. 
We will return to this quantity in our future studies.

The comparison to the purely perturbative curve shows qualitative 
agreement. A meaningful quantitative comparison requires a careful
study of lattice artifacts which may be quite noticeable in the
region of small $r$, where perturbation theory applies.
Indeed perturbation theory by itself suggests that at least
$r\leq\frac12 r_0$ is necessary\cite{Necco:2001gh}, in particular
when the new 4-loop beta function is taken into account as discussed
in appendix B.

\section{Conclusions}

We have presented a detailed analysis of the static potential defined by
the HYP2 action for the static quarks.
\fig{f_V} and \fig{f_c} show the quality of our
data. Judged by a comparison of HYP and HYP2, cut-off effects in the 
potential appear to be small.
The scale $r_0/a$ can be determined with precision better than 1\%. 
We observe large
effects due to dynamical fermions in the quantity $c(r)$ defined in \eq{ccoeff}.

As can be seen in Table \ref{t_force_c} the error on the force grows faster with
the distance $r$ as compared to the pure gauge case (see Table 2 of
\cite{Luscher:2002qv}). This effect is amplified by $r^3$ for the quantity $c(r)$.
It remains to be seen whether the inclusion of fermionic correlators in the 
variational
basis will lead to an improvement due to a larger overlap with the ground state
and the resulting earlier start of a plateau.

A precise study of the static potential is relevant for phenomenology
in an indirect but important way. As reviewed in \cite{QWG}, 
there is an impressive effort to apply
potential non-relativistic QCD (pNRQCD)\cite{PNRQCD} 
to the top -- anti-top production in a
future $e^+\;e^-$ collider and to many other processes. 
This effective theory includes ultrasoft gluons and is treated 
perturbatively in the QCD coupling. While the potential of pNRQCD 
is not the same
as the static potential, the two are intimately related; they differ only 
starting at NNNLO accuracy. It is hence
very useful to understand where the perturbative approximation to 
the static potential can be trusted. 
\fig{f_c} is a start for that, but a precise investigation
requires the removal of lattice artifacts \cite{Necco:2001xg}. 
In the future we plan to work 
both on this connection to the perturbative regime of QCD
and on the large distance, string breaking, region.

\bigskip

{\bf Acknowledgement.}
We are grateful to Nazario Tantalo for extensive checks of the Wilson loop
measurements and to Stefan Schaefer for help in checking the HYP smearing.
We thank Nikos Irges for discussions on the quantity $c(r)$ and 
Valentina Forini for discussions on the AdS/CFT correspondence.
We further thank NIC and
the Zuse Institute Berlin for allocating computing
resources to this project. Part of the Wilson loop measurements were performed
on the PC-cluster of DESY, Zeuthen.

\begin{appendix}
\section{Improvement \label{s_appa}}
%
\begin{table}[p!]
 \centering
\begin{tabular}{ccccc}
\hline\\
 $r/a$  & \multicolumn{2}{c}{$\rI/a$} &  \multicolumn{2}{c}{$\tilde{r}/a$} 
\\[0.5ex]
   & HYP & HYP2 & HYP & HYP2 \\[1.0ex] 
\hline\\
 4 & 3.48560 & 3.55805 & 3.97292 & 4.04631\\
 5 & 4.45369 & 4.52674 & 4.93158 & 5.02609\\
 6 & 5.44414 & 5.50073 & 5.91700 & 5.99970\\
 7 & 6.44353 & 6.48362 & 6.91468 & 6.97787\\
 8 & 7.44614 & 7.47397 & 7.91743 & 7.96350\\
 9 & 8.44969 & 8.46922 & 8.92199 & 8.95537\\
10 & 9.45331 & 9.46734 & 9.92696 & 9.95146\\
11 & 10.4567 & 10.4670 & 10.9318 & 10.9501\\
12 & 11.4598 & 11.4676 & 11.9362 & 11.9503\\
13 & 12.4625 & 12.4685 & 12.9403 & 12.9512\\
14 & 13.4649 & 13.4697 & 13.9440 & 13.9527\\
15 & 14.4671 & 14.4709 & 14.9472 & 14.9543\\
16 & 15.4690 & 15.4721 & 15.9502 & 15.9560\\
17 & 16.4707 & 16.4733 & 16.9529 & 16.9576\\
18 & 17.4723 & 17.4745 & 17.9553 & 17.9593\\
19 & 18.4737 & 18.4755 & 18.9575 & 18.9609\\
20 & 19.4750 & 19.4765 & 19.9595 & 19.9624\\
21 & 20.4762 & 20.4775 & 20.9613 & 20.9638\\
22 & 21.4772 & 21.4784 & 21.9630 & 21.9651\\
23 & 22.4782 & 22.4792 & 22.9645 & 22.9664\\
24 & 23.4791 & 23.4800 & 23.9659 & 23.9676\\
25 & 24.4799 & 24.4807 & 24.9672 & 24.9687\\
26 & 25.4807 & 25.4814 & 25.9685 & 25.9698\\
27 & 26.4814 & 26.4820 & 26.9696 & 26.9708\\
28 & 27.4820 & 27.4826 & 27.9706 & 27.9717\\
29 & 28.4827 & 28.4831 & 28.9716 & 28.9726\\
30 & 29.4832 & 29.4837 & 29.9726 & 29.9734\\
31 & 30.4838 & 30.4842 & 30.9734 & 30.9742\\
32 & 31.4843 & 31.4846 & 31.9742 & 31.9749\\[1.0ex]
\hline
\end{tabular}
 \caption{\small The values of the improved distances $\rI/a$ \eq{rIeq} and
   $\rt/a$ \eq{rteq} extrapolated to $L/a\to\infty$ for the case of HYP and
   HYP2 smearings. We show 6 significant digits for all values of $r/a$,
   where the last digit is rounded.
}
 \label{t_rimpr}
\end{table}
%

The tree level perturbative expression for the
static potential, which is extracted
from Wilson loops where the static quark line is HYP smeared, is given in
\cite{Hasenfratz:2001tw,rolanddipl}.
From it we easily derive the formula for $\rI$ defined from \eq{rIdef}:
\bea
(4\pi\rI^2)^{-1} & = & -[G_{\rm HYP}(r,0,0)-G_{\rm HYP}(r-a,0,0)]/a \,,
\label{rIeq}
\eea
with
\bea
G_{\rm HYP}(\vec{r}) & = & \frac{1}{a}\int_{-\pi}^\pi
\frac{{\rm d}^3p}{(2\pi)^3}
\frac{\prod_{j=1}^3\cos(x_jp_j/a) \times 
f_{\rm sm}(p)}{\sum_{j=1}^3\hat{p}_j^2} \,, \label{GHYP}
\eea
where $\vec{r}=(x_1,x_2,x_3)$, $\hat{p}_j=2\sin(p_j/2)$ and the smearing
factor is
\bea
f_{\rm sm}(p) & = &
\left[1-\frac{\alpha_1}{6}\sum_{j=1}^3\hat{p}_j^2\Omega_{j0}(p)\right]^2 \\
\Omega_{j0}(p) & = & 1 + \alpha_2(1+\alpha_3) -
\frac{\alpha_2}{4}(1+2\alpha_3)(\hat{p}_1^2+\hat{p}_2^2+\hat{p}_3^2-\hat{p}_j^2)
+ \frac{\alpha_2\alpha_3}{4}\prod_{\tau\neq0,j}\hat{p}_\tau^2 \nonumber \\
\eea
($f_{\rm sm}=1$ for unsmeared static quark lines).

The distance $\rt$ defined from \eq{rtdef} is given
in the case of HYP smeared static quarks by
\bea
\rt^{-3} & = & 2\pi[G_{\rm HYP}(r+a)+G_{\rm HYP}(r-a)-2G_{\rm HYP}(r)]/a^2 \,.
\label{rteq}
\eea
In practice we evaluate the momentum integral in \eq{GHYP} by discrete
momentum sums over $p_j=2\pi n_ja/L$, $n_j=0,1,\ldots,L/a-1$. In
\tab{t_rimpr} we quote the results for $\rI/a$ and $\rt/a$ obtained from
extrapolations $L/a\to\infty$. The latter are done with the method explained
in Appendix D of \cite{Bode:1999sm} and we consider lattice sizes larger than
$L/a=128$ up to $L/a=512$. Due to the symmetry under $p_j\to-p_j$ of the
integrand only odd powers of $a/L$ can appear in the expansion in powers of 
$a/L$ and in general this evaluation of the integral is the 
application of a trapezoidal rule, which has discretization errors
of order $(a/L)^2$. Thus the leading correction is $s_1(a/L)^3$. 
The data for $\rI/a$ and $\rt/a$ are very well 
fitted by a polynomial $s_0+s_1(a/L)^3+s_2(a/L)^5$ and 
we added terms $s_3(a/L)^7+s_4(a/L)^9$ to estimate the systematic
error of the extrapolations. In \tab{t_rimpr} we list
the extrapolated values with six significant digits. 

\section{Perturbation theory for $c(r)$ \label{s_appb}}

We consider QCD with $\Nf$ massless dynamical quark flavors.
The quantity $c(r)$ in \eq{ccoeff} defines a renormalized coupling ($\CF=4/3$), 
\bea
 \gc^2(\mu) = - {4\pi \over \CF} c(r)\,,\;\mu=1/r \,.
\eea
It is very similar to $\gqq^2(\mu) = {4\pi\over \CF} r^2 F(r)\,,\;\mu=1/r$
discussed in \cite{Necco:2001gh}. The relation is
\bea
 \gc^2 & = & \gqq^2 + \gqq\beta_\mathrm{qq} \,.
\eea
For a perturbative evaluation of one-scale quantities
such as $c(r)$ it is natural\footnote{It has also been observed in more than one case
that it also yields a good perturbative description of the non-perturbative 
behavior.} to just integrate the renormalization group equation
\bea
  \mu\frac{{\rm d}}{{\rm d}\mu}\gc(\mu) & = & \betac(\gc(\mu))\,.
\eea
We do this in the precise form of
\bea
 {\Lambda_\mathrm{c} \over \mu}  &=& 
  \left(b_0\gc^2\right)^{-b_1/(2b_0^2)} {\rm e}^{-1/(2b_0\gc^2)}
           \exp \left\{-\int_0^{\gc} {\rm d}  x
          \left[\frac{1}{ \betac(x)}+\frac{1}{b_0x^3}-\frac{b_1}{b_0^2x}
          \right]
          \right\} \,, 
\nonumber \\  \label{e_lambdapar}
\eea
where for $\betac$ the truncated perturbative expansion is inserted,
but the integral is (numerically) evaluated as it stands. 
The Lambda-parameter in the c-scheme is 
\bea
 \Lambda_c & = & {\rm e}^{-1/2}\Lambda_\mathrm{qq} = 
 \Lambda_{\MSb}{\rm e}^{k_1/(8\pi b_0) \,-\,1/2} \,,
\eea
where $k_1=\frac{1}{4\pi}(a_1+a_2\Nf)$, $a_1=-35/3+22\gE$ and 
$a_2=2/9-4\gE/3$ \cite{Necco:2003jf}\,.
We now turn to the perturbative  beta function.

\subsection{Perturbative beta function in the c-scheme}

The expansion of the potential in the $\MSb$ coupling is now
known to a high accuracy. After the $\gms^6$ term
\cite{pot:2loop1,pot:2loop2}, the resummation of the infrared
divergent diagrams appearing first at the next order was 
performed\cite{Brambilla:1999qa,PNRQCD}, yielding a
$\sim \gms^8\log(\gms^2)$ term. Recently also
the full three-loop computation was finished by two groups  
\cite{Smirnov:2009fh,Anzai:2009tm}.
Due to the $\gms^8\log(\gms^2)$ term in the 
potential\cite{Brambilla:1999qa,PNRQCD}, 
the beta function has a perturbative expansion
\bea
\betac(\gc) & = &
-\gc^3 [\sum_{n=0}^3 \bc_n \gc^{2n} +  \bc_{3,l} \gc^{6}\log(\CA \gc^2/(8\pi)) + \mathrm{O}({\gc^8})] \,,
\label{betapt}
\eea
with the universal coefficients ($\CA=3$)
\bea
\bc_0 = b_0 & = & \frac{1}{(4\pi)^2}(11\CA/3-2\Nf/3)\,, \label{b0} \\
\bc_1 = b_1 & = & \frac{1}{(4\pi)^4}(34\CA^2/3-10\CA\Nf/3-2\CF\Nf) \,. \label{b1}
\eea

We now describe how the non-universal coefficients are obtained
from the results in the literature. 
Our starting point is Eq. (40) of \cite{Brambilla:2009bi}, which is the
expansion of the static potential $V(r)$, denoted ``static energy''
in \cite{Brambilla:2009bi}, in the $\MSb$ strong coupling 
$\as=\gms^2(1/r)/(4\pi)$ 
derived from the above mentioned work. Introducing the notation
$V(r)=-\CF G(\as)/r$, we obtain an expansion for $\ac = \gc^2/(4\pi)$ :
\bea
\ac & = & \frac{1}{2}\,r^2\,G^{\prime\prime}(\as) 
           - r\,G^{\prime}(\as) + G(\as)  \\
& = & \as + d_1\,\as^2 + d_2\,\as^3 + d_3\,\as^4 
+ d_{3,l}\,\as^4\,\ln\left(\frac{\CA\as}{2}\right) 
+ \mathcal{O}(\as^5)\,, \label{cexpansion}
\eea
where the primes in the first equation mean derivatives with respect to $r$
and the coefficients of the expansion are
\bea
d_1 & = & \frac{1}{4\pi}\left(\tilde{a}_1- 3 b_0 (4\pi)^2\right) \,,\\
d_2 & = & \frac{1}{(4\pi)^2}
\left(\tilde{a}_{2,s} + 4 b_0^2 (4\pi)^4 - 3 b_1 (4\pi)^4 - 6 \tilde{a}_1 b_0 (4\pi)^2 \right) \,\\
d_3 & = &
\frac{1}{(4\pi)^3}\left(
a_3 + 12 \tilde{a}_1 b_0^2 (4\pi)^4 - 6 \tilde{a}_1 b_1 (4\pi)^4 +
10 b_0 b_1 (4\pi)^6  \right. \nonumber \\
& & \left. - 3 b_2 (4\pi)^6 - 9 \tilde{a}_{2,s} b_0 (4\pi)^2  \right) \,, \\
d_{3,l} & = & \frac{\CA^3}{12\pi} \,.
\eea
The coefficients $b_2$ and
$b_3$  of the beta function in
the $\MSb$ scheme,
$\beta_{\rm\overline{MS}}(\gms)  \sim  -\gms^3\sum_{n\geq0}\gms^{2n} b_n$,
can be found in \cite{vanRitbergen:1997va,Czakon:2004bu}. The
coefficient $\tilde{a}_1$ is defined in Eq. (7) and the coefficient
$\tilde{a}_{2,s}$ in Eq. (8) of \cite{Brambilla:2009bi}, they both
depend on $\Nf$. The coefficient $a_3$ is
\bea
a_3 & = &
\frac{4^4}{3\CF}\left( c_0(\Nf) + 2 \gE c_1(\Nf) + 
(4 \gE^2 + \pi^2/3) c_2(\Nf) \right. \nonumber \\
& & \left. + (8 \gE^3 + 2 \pi^2 \gE + 16 \zeta(3)) c_3(\Nf) \right)
\eea
where
\bea
c_0(\Nf) & = & 219.59 + \left(a_3^{(1)}\Nf + a_3^{(2)}\Nf^2 +
a_3^{(3)}\Nf^3\right)/4^3 \,.
\eea
From \cite{Brambilla:2009bi, Brambilla:2010pp} we get\footnote{We thank the
authors of \cite{Brambilla:2009bi} for communication on the value of
$c_0(0)$.} $c_0(0)$ and
the coefficients $a_3^{(1)}$, $a_3^{(2)}$ and $a_3^{(3)}$ are given in
Eq. (6) of \cite{Smirnov:2008pn}.
The coefficients $c_1(\Nf)$, $c_2(\Nf)$ and $c_3(\Nf)$ are defined
in Eqs. (10), (11) and (12) of \cite{Chishtie:2001mf} respectively.
%
\begin{figure}[t]
 \begin{center}
     \includegraphics[width=11cm]{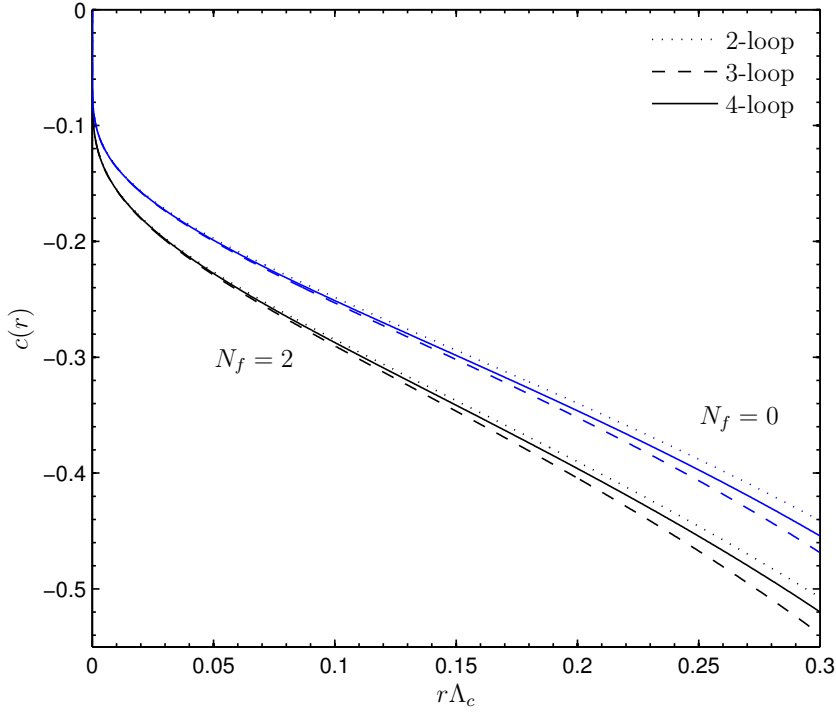}
 \end{center}
 \vspace{-0.5cm}
 \caption{\small The perturbative running of the quantity $c(r)$ obtained from
   \eq{e_lambdapar} using the 2-loop (dotted lines), 3-loop (dashed lines) and
   4-loop (continued lines) beta function $\betac(\gc)$ for $\Nf=0,2$.}
 \label{f_c_pt}
\end{figure}
%

The non-universal coefficients $\bc_2$ and $\bc_3$ as
well as the coefficient $\bc_{3,l}$ may now be computed
by differentiating $\betac = {2\pi\over \gc} \mu {{\rm d}\ac \over {\rm d}\mu}$
with $\ac$ of \eq{cexpansion}, where the
$\overline{\rm MS}$ beta function is used. This first yields
$\betac$ as a function of $\as$ from which we change to
$\betac(\gc)$ by inserting the 
inverted \eq{cexpansion}, 
$\as =\ac + \ldots -  d_{3,l}\,\ac^4\,\ln\left(\frac{\CA\ac}{2}\right)$.

Carrying this out in MAPLE we find
\bea
\bc_2 & = & b_2 - 5 b_0^3 + \tilde{a}_{2,s} b_0 (4\pi)^{-4} - \tilde{a}_1 b_1 (4\pi)^{-2} 
- \tilde{a}_1^2 b_0 (4\pi)^{-4} \label{bc2}\\
 &=& (4\pi)^{-3} [0.98165 -  0.16738\Nf - 0.00212 \Nf^2 + 0.00026 \Nf^3]  \nonumber \\
\bc_3 & = &
b_3 - 2 \tilde{a}_1 b_2 (4\pi)^{-2} + 2 a_3 b_0 (4\pi)^{-6} 
+ \frac{1}{3} \CA^3 b_0 (4\pi)^{-4} - 25 b_0^2 b_1 \nonumber \\
& & - 6 \tilde{a}_1 \tilde{a}_{2,s} b_0 (4\pi)^{-6} + \tilde{a}_1^2 b_1 (4\pi)^{-4}
- 36 b_0^4 + 4 \tilde{a}_1^3 b_0 (4\pi)^{-6}  \label{bc3} \\
 &=& (4\pi)^{-4} [0.12206 + 0.09696 \Nf - 0.01899 \Nf^2 + 0.0004458 \Nf^3
\nonumber \\
&&                + 0.0000195 \Nf^4] \nonumber \\
\bc_{3,l} & = & \frac{2}{3} \CA^3 b_0 (4\pi)^{-4} = (4\pi)^{-4} 
    [1.25385 - 0.07599\Nf] \,. \label{bc3l}
\eea
As in the $\MSb$ scheme the coefficients $(4\pi)^{n+1} b_n$ are of order one
and thus the beta fuction has a well-behaved expansion up to
couplings $\ac$ of the order of $1/3$. The perturbative running of $c(r)$
is shown in \fig{f_c_pt}.

The "asymptotic convergence" of the series \eq{cexpansion} is not good.
It can be substantially improved by matching the couplings at a different 
scale, i.e., by expressing $\ac(s/r)$ as a function of $\as(1/r)$ and choosing
$s=s_0=\Lambda_c/\Lambda_{\MSb}$ (``fastest apparent convergence'',
cf. \cite{Necco:2001gh,Necco:2003jf}). The resulting curves for $\ac$ are
hardly distinguishable from the ones shown in \fig{f_c_pt}.

\subsection{Perturbative beta function in the qq-scheme}

In the same way one obtains the beta function in the qq-scheme. We update the 
formulae given in \cite{Necco:2003jf} to include the 4-loop term:
\bea
\bqq_2  &=& (4\pi)^{-3} [1.6524 - 0.28933 \Nf + 0.00527 \Nf^2 + 0.00011 \Nf^3]  \\
\bqq_3  &=& (4\pi)^{-4} [4.94522 - 1.07965 \Nf + 0.079107 \Nf^2 - 0.002774 \Nf^3
\nonumber \\
&&                + 0.000051 \Nf^4] \\
\bqq_{3,l} & = & \frac{2}{3} \CA^3 b_0 (4\pi)^{-4} = (4\pi)^{-4} 
    [1.25385 - 0.07599\Nf] \,.
\eea
Perturbation theory in the c-scheme appears much better 
behaved than in the qq-scheme. Since this can only be considered an
accident we come to the same conclusion as \cite{Necco:2003jf}, namely that 
the perturbative description of the static potential is accurately valid only 
at rather small values of $r$, where $\aqq(1/r)\approx1/4$. Unfortunately
these distances are close to present lattice spacings.
In particular the data presented in this paper are not good enough to extract 
the $\Lambda$ parameter through \eq{e_lambdapar} or variants thereof.

\end{appendix}

\bibliography{potbiblio}           
\bibliographystyle{h-elsevier}   

\end{document}